\documentclass[aps]{revtex4}

\usepackage{amssymb}
\usepackage{amsmath}
\usepackage{graphicx}
\usepackage{graphics}
\usepackage{color}
\usepackage{epstopdf}
\usepackage{hyperref} 
\usepackage{setspace}

\begin{document}

\newcommand{\blue}[1]{\textcolor{blue}{#1}}
\newcommand{\new}{\blue}
\newcommand{\green}[1]{\textcolor{green}{#1}}
\newcommand{\modif}{\green}
\newcommand{\red}[1]{\textcolor{red}{#1}}
\newcommand{\attention}{\red}

\setstretch{1.25}

\title{Chiral Symmetry in the Confinement Phase of QCD}

\author{S. D. Campos} \email{sergiodc@ufscar.br}
\affiliation{Applied Mathematics Laboratory-CCTS/DFQM, Federal University of \\ S\~ao Carlos, Sorocaba, S\~ao Paulo CEP 18052780, Brazil}

\date{\today}



\begin{abstract}
Based on the Pomeranchuk theorem, one constructs the $\delta(s)$ parameter to measure the difference between experimental data for the particle-particle and particle-antiparticle total cross section at same energy. The experimental data for the proton-proton and proton-antiproton total cross section were used to show that, at the same energy, this parameter tends to zero as the collision energy grows. Furthermore, one assumes a classical description for the total cross section, dividing it into a finite number of non-interacting disjoint cells, each one containing a quark-antiquark pair subject to the confinement potential. Near the minimum of the total cross section, one associates $\delta(s)$ with the entropy generated by these cells, analogously to the XY-model. Using both the Quigg-Rosner and Cornell confinement potentials and neglecting other energy contributions, one can calculate the internal energy of the hadron. One obtains that both the entropy and internal energy possess the same logarithmic dependence on the spatial separation between the pairs in the cell. The Helmholtz free energy is used to estimate the transition temperature, which is far from the temperature widely related to the Quark-Gluon Plasma. 

\end{abstract}


\maketitle


\section{Introduction}

The so-called forward quantities ($t=0$) play a fundamental role in high energy scattering physics. As well-known, the precise measurement of the total cross section $\sigma_{tot}(s)$, as well the $\rho(s)$-parameter and the slope $B(s,t)$, can help the normalization and calibration of beams as well detectors. Furthermore, the optical theorem connects $\sigma_{tot}(s)$ with the imaginary part of the forward scattering amplitude and, through the well-known derivative dispersion relations, one can obtain the full-forward scattering amplitude. Hereafter, $s$ and $t<0$ are, respectively, the squared energy and the squared momentum transfer in the center-of-mass system. 

In Quantum Mechanics (QM), $\sigma_{tot}(s)$ is interpreted as the probability of occurrence of an event in a collision process, representing the sum over all possible final states. In potential theory, the total cross section is determined uniquely from a given potential. Classically, on the other hand, $\sigma_{tot}(s)$ can be viewed as representing the {\it effective} area in a head-on collision. Necessary to emphasize that the result of QM is four times the classical, which has been interpreted as if the wave associated with the scattering takes into account the area of the sphere \cite{D.R.Bes.Quantum.Mchanics.A.Modern.and.Concise.Introductory.Course.Third.Edition.Springer.2012}. 

There are some cases in physics for which physical observables are interpreted in terms of an effective area. An ordinary example is work done by an external force. A further example, much more sophisticated, is given by the entropy-area law for the black holes \cite{S.W.Hawking.Phys.Rev.Lett.26.1344.1971,S.W.Hawking.Comm.Math.Phys.25.152.1972,J.D.Bekenstein.Lett.Nuovo.Cim.4.737.1972,J.D.Bekenstein.Phys.Rev.7.2333.1973}. Even in the definition of confinement (definitively a not easy task) in Quantum Chromodynamics (QCD), one can associate the planar Wilson loops with an area-law falloff \cite{J.Greensite.K.Matsuyama.Phys.Rev.D96.094510.2017}. Also, the concept of area is closely related to the study of phase transitions. 

In solid-state physics, the typical way to introduce the study of phase transition is using 2D-systems and looking for the emergence of some physical modification in the system taking into account a certain critical value (temperature is the usual parameter). The result of such analyzes has culminated in the study of the physical properties of the so-called XY-model, intensively studied years ago \cite{H.E.Stanley.T.A.Kaplan.Phys.Rev.Lett.17.913.1966}. In particular, the XY-model seemed to violate the Mermin-Wagner-Hohenberg-Coleman theorem \cite{N.D.Mermin.H.Wagner.Phys.Rev.Lett.17.1133.1966,P.C.Hohenberg.Phys.Rev.158.383.1967,S.Coleman.Commun.Math.Phys.31.259.1973}, i.e. considering a system with short-range interactions and dimension $d\leq 2$, then there is no symmetry breaking (at finite temperatures) of continuous symmetries. However, the study carried out by Berezinskii \cite{V.L.Berezinskii.Zh.Eksp.Teor.Fiz.59.907.1970} and, independently, by Kosterlitz and Thouless \cite{J.M.Kosterlitz.D.J.Thouless.J.Phys.C6.1181.1973} revealed the role played by topological excitations to explain the divergences present in the XY-model. Nowadays, this effect is known as the Berezinskii-Kosterlitz-Thouless (BKT) phase transition, occurring in two-dimensional systems and possessing a topological character. By a topological phase transition, it should be understood that the physical space of the system has a topology below
the critical value and another topology above this value. Originally, a system where the BKT phase transition takes place is characterized by the emergence of unbound vortices, i.e. the arising of single vortices is energetically favored over the vortex-antivortex pair \cite{V.L.Berezinskii.Zh.Eksp.Teor.Fiz.59.907.1970,J.M.Kosterlitz.D.J.Thouless.J.Phys.C6.1181.1973}. This effect is only possible once the entropy controls the free energy of the system, above the phase transition.  

In the context of particle physics, the general belief attaches to the Hagedorn temperature \cite{r.hagedorn_nuovo_cim_supp_3_147_1965,r.hagedorn_nuovo_cim_56a_1027_1968} the phase transition from the hadronic matter to Quark-Gluon Plasma (QGP) \cite{J.C.Collins.M.J.Perry.Phys.Rev.Lett.34.1353.1975,N.Cabibbo.G.Parisi.Phys.Lett.B.59.67.1975,E.V.Shuryak.Sov.Phys.JETP.47.212.1978} (part of the vast literature about QGP can be found in Refs. \cite{E.V.Shuryak.arXiv:1812.01509v2.[hep-ph].2018,L.S.Kisslinger.Int.J.Mod.Phys.A.32.1730008.2017}, and references therein). This is a dynamical phase transition representing the spontaneous breaking of the chiral symmetry, being a formal prediction of QCD.  


Based on the above discussion, the first question arises: is there some topological phase transition in high energy scattering? To answer this question, it is necessary first to look at the experimental data for proton-proton ($pp$) and proton-antiproton ($p\bar{p}$) total cross sections. These data exhibit a minimum for $\sigma_{tot}(s)$ at some collision energy. As well-known, for a long time, the general belief was the total cross section should be asymptotically constant. Only the measurement of Serpukhov on $\pi^\pm p$ and $K^\pm p$, and later ISR and FNAL on $pp$ and $p\bar{p}$ total cross sections, had shown the rise of $\sigma_{tot}(s)$. Thus, a possible candidate where the phase transition takes place may be the minimum of $\sigma_{tot}(s)$. To produce a topological phase transition in the same way as BKT, it is necessary to introduce a second question: may the classical view of the total cross section be connected with the entropy of the hadron during the collision process? If this assumption is valid, then the next goal is to obtain the internal energy of the hadron.  

To show that the entropy may be obtained from $\sigma_{tot}(s)$, one uses the Pomeranchuk theorem \cite{I.Ia.Pomeranchuk.Sov.Phys.JETP.7.499.1958} to construct the $\delta(s)$ parameter, relating $pp$ and $p\bar{p}$ total cross sections. Considering the classical description of the total cross section, one divides $\sigma_{tot}(s)$ into a finite number of non-interacting disjoint cells, each one containing a quark-antiquark ($q\bar{q}$) pair. Thus, one obtains a quantity that can be related to the entropy of the system. This entropy depends on the logarithm of the ratio of the total cross section to the size of the cell.  

The next step is to calculate the hadron internal energy from the confinement potential. One uses here two confinement potentials: Quigg-Rosner \cite{C.Quigg.J.L.Rosner.Phys.Lett.B71.153.1977,C.Quigg.J.L.Rosner.Phys.Rep.56(4).167.1979} and Cornell \cite{e_eichten_Phys_Rev_Lett_34_369_1975,e_eichten_Phys.Rev.D17.3090.1978,e_eichten_Phys.Rev.D21.203.1980}. As shall be seen, the entropy and internal energy in the system possess the same logarithmic dependence, competing by the Helmholtz free energy and resulting in a topological phase transition, exactly as the BKT phase transition \cite{V.L.Berezinskii.Zh.Eksp.Teor.Fiz.59.907.1970,J.M.Kosterlitz.D.J.Thouless.J.Phys.C6.1181.1973}. 

It is important to stress that the entropy calculation proposed here is not the same one measured in, for example, multiparticle systems created in high-energy collision events \cite{A.Bialas.W.Czyz.J.Wosiek.Acta.Phys.Polon.30.107.1999,A.Bialas.W.Czyz.Phys.Rev.D61.074021.2000,A.Bialas.W.Czyz.ActaPhys.Polon.31.687.2000,K.Fialkowski.R.Wit.Phys.Rev.D62.114016.2000}. The main point in this work is to study the effects of the entropy by analyzing the hadronic total cross section as the energy grows, trying to understand, from the thermodynamic point of view, the effects of the confinement of quarks and gluons. A possible interpretation for such phase transition may be the chiral symmetry restoration \cite{L.McLerran.R.D.Pisarski.Nucl.Phys.A796.83.2007,L.YA.Glozman.R.F.Wagenbrunn.Mod.Phys.Lett.A23.2385.2008} in the confinement phase of QCD. In this scenario, the emergence of free quarks above the phase transition can occur, resulting in an effect quite similar to the one at XY-model. 

This work has been organized as follows. In Section \ref{sigma}, one introduces the $\delta(s)$ parameter, analyzing its physical meaning based on the experimental data for $pp$ and $p\bar{p}$ total cross section. Section \ref{bkt} presents the BKT phase transition according to the two confinement potentials adopted. A discussion of the results is left to the final section \ref{fr}.

\section{\label{sigma}The $\delta(s)$ Parameter}

At the end of the 1950s, Pomeranchuk has suggested that particle-particle and particle-antiparticle total cross sections tend to be asymptotically equal at $s\rightarrow\infty$ \cite{I.Ia.Pomeranchuk.Sov.Phys.JETP.7.499.1958}. Thus, one can write for the asymptotic condition $s\rightarrow\infty$
\begin{eqnarray}\label{eq:asym_1}
	\frac{\sigma_{tot}^{p\bar{p}}(s)}{\sigma_{tot}^{pp}(s)}\rightarrow 1,
\end{eqnarray} 

\noindent which is a result obtained by Grunberg and Truong considering that $\sigma_{tot}(s)$ grows with energy \cite{C.Grunberg.T.N.Truong.Phys.Rev.Lett.31.63.1973}. At high $s$, the pomeron is the exchange particle responsible by (\ref{eq:asym_1}) and should have the vacuum quantum numbers \cite{L.F.Foldy.R.F.Peierls.Phys.Rev.130.1585.1963}. Thus, the pomeron cannot distinguish particle from antiparticle, explaining why the total cross sections tend to the same value as stated in (\ref{eq:asym_1}). At low $s$, a sub-leading term, the so-called odderon, becomes dominant. That particle is defined as the leading singularity in the complex plane contributing to the odd crossing amplitude.  In this way, it predicts the differences in the differential cross section for $pp$ and $p\bar{p}$. In other words, this particle exchange has different couplings to particle and antiparticle, resulting in that $pp$ and $p\bar{p}$ total cross sections have different slopes.

Recently, a possible experimental evidence for the odderon was obtained \cite{G.Antchev.etal.TOTEM.Coll.Eur.Phys.J.C79.785.2019}, being subject to discussions in the literature \cite{E.Martynov.B.Nicolescu.Phys.Lett.B778.414.2018,V.A.Khoze.A.D.Martin.M.G.Ryskin.Phys.Lett.B780.352.2018,A.Szczurek.P.Lebiedowicz.PoS.DIS2019.071.2019}. 

Based on the above theoretical result, one proposes the use of the following parameter defined below 
\begin{eqnarray}\label{eq:asym_2}
	\delta(s)=\ln\frac{\sigma_{tot}^{p\bar{p}}(s)}{\sigma_{tot}^{pp}(s)},
\end{eqnarray} 

\noindent as a possible way to measure the difference between both the $pp$ and $p\bar{p}$ experimental data set \textit{at same energy}. Note that if the Pomeranchuk theorem (\ref{eq:asym_1}) is satisfied, then $\delta(s)\rightarrow 0$ as $s\rightarrow\infty$. 

\begin{figure}
	\centering{
		\includegraphics[scale=0.35]{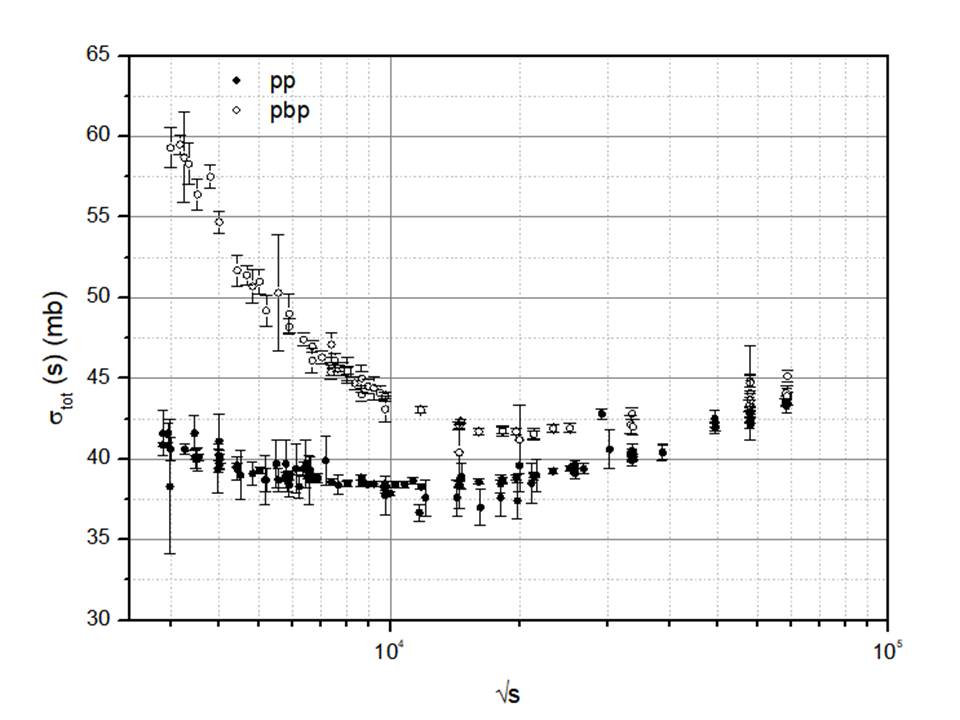}
		\caption{\label{fig:fig_sigma} Total cross section experimental data in the energy range $3.5$ GeV $<\sqrt{s}< 63.0$ GeV, collected from the Particle Data Group \cite{PDG-PhysRev-D98-030001-2018}. The experimental data here are not necessarily at the same energy, being used only as a guide to the reader.}}
\end{figure}

\begin{figure}
	\centering{
		\includegraphics[scale=0.35]{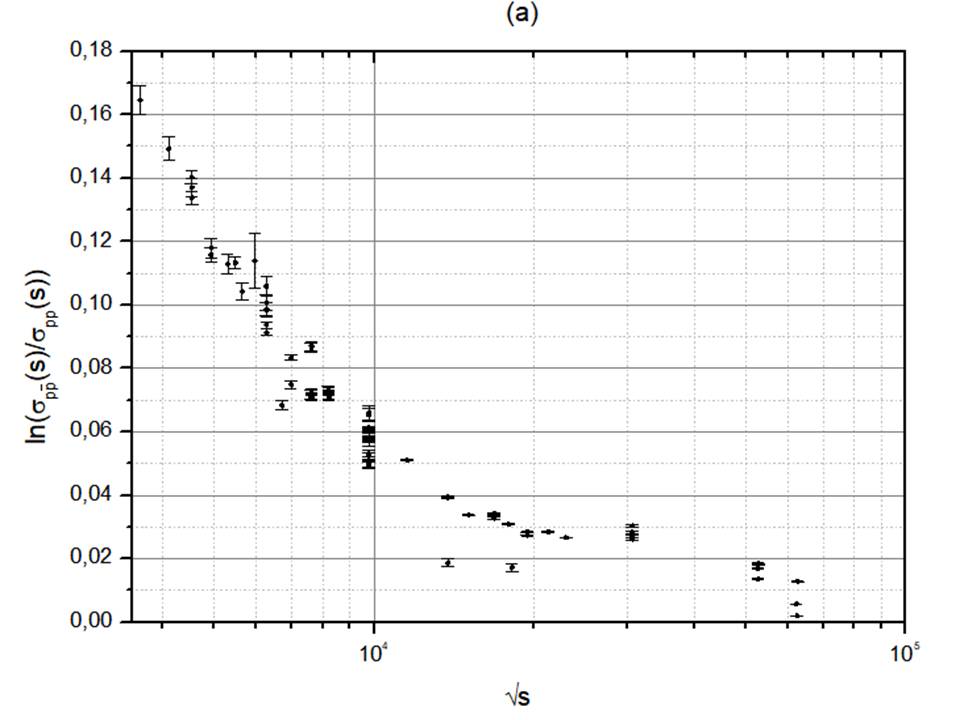}
		\includegraphics[scale=0.35]{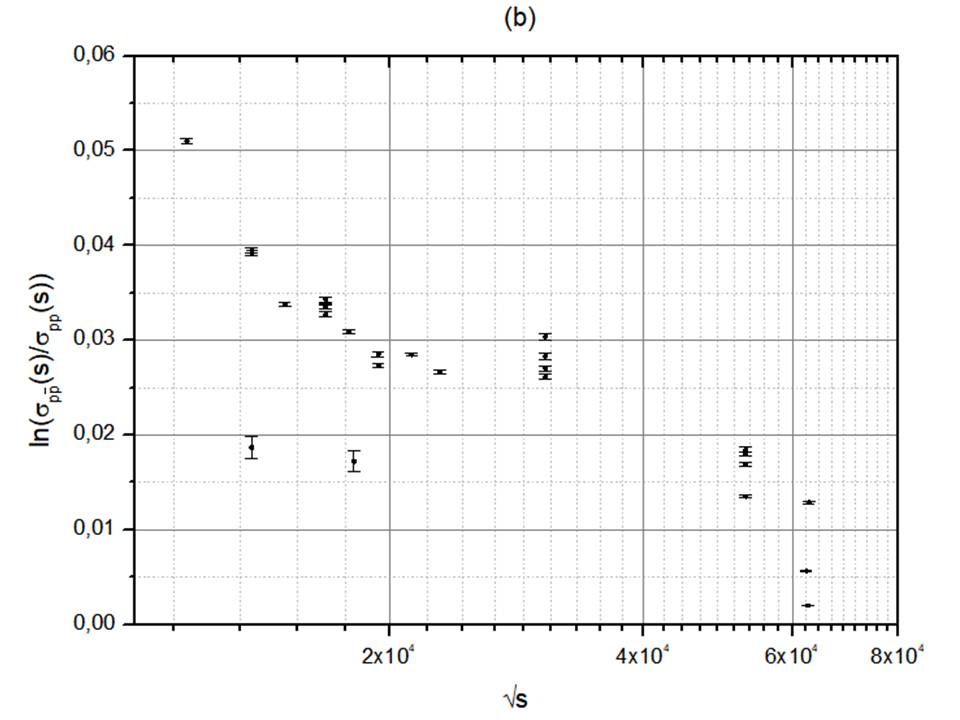}
		\caption{\label{fig:fig_1}The $\delta(s)$ parameter behavior depending on $pp$ and $p\bar{p}$ experimental data. In panel (a), one has all the experimental data at same energy for $pp$ and $p\bar{p}$. Panel (b) shows the shoulder near the minimum of the total cross sections, shown in Figure \ref{fig:fig_sigma}. Apparently, $\delta(s)$ tends to zero as $s\rightarrow\infty$.}}
\end{figure}

Figure \ref{fig:fig_sigma} shows all the available experimental data for $pp$ and $p\bar{p}$ total cross section in the energy range $3.5$ GeV $<\sqrt{s}<63.0$ GeV. The data were collected from the Particle Data Group \cite{PDG-PhysRev-D98-030001-2018}, and there was any data selection. This figure is used only as a guide to the analysis of $\delta(s)$, given below. 

To produce the values observed in Figure \ref{fig:fig_1} for $\delta(s)$, one uses only experimental data for $pp$ and $p\bar{p}$ {\it at the same energy} in both panels. Unfortunately, there is no experimental data for both total cross sections \textit{at the same energy} above $63.0$ GeV. Thus, for practical uses, the high energy limit is limited here to the medium energy range. The error propagation is performed as usual
\begin{eqnarray}\label{eq:error}
	\Delta(\delta)=\delta\sqrt{\frac{\Delta(\delta_1)^2}{\delta_1^2}+\frac{\Delta(\delta_2)^2}{\delta_2^2}}.
\end{eqnarray}

In Fig. \ref{fig:fig_1}a, one has the global behavior of $\delta(s)$ based on the experimental data collected. It is interesting to note that, despite the collision energy $s$ is far from the so-called asymptotia, the value of the logarithmic ratio (\ref{eq:asym_2}) at $\sqrt{s}=62.7$ GeV is very close to zero, $\delta(s)=0.0128\pm0.0001$. Notice that $\delta(s)$ grows as the collision energy diminishes toward the Coulomb-Nuclear region possibly indicating that the role played by the odderon exchange is more evident when contributions from secondary reggeons (pomeron) can be neglected.

In Fig. \ref{fig:fig_1}b, one observes the small shoulder in the energy range $10.0$ GeV $\lesssim\sqrt{s}\lesssim30.0$ GeV. Its occurrence may be explained as follows. First of all, it is important to stress that in the energy range $10.0$ GeV $<\sqrt{s}<30.0$ GeV, one observes the minimum of $\sigma_{tot}(s)$ (for $pp$ and $p\bar{p}$), as can be viewed in Figure \ref{fig:fig_sigma}. The minimum, however, does not necessarily occur at the same energy. Thus, one may have two minima and, further, in this energy range $\sigma_{tot}^{pp}(s)<\sigma_{tot}^{p\bar{p}}(s)\rightarrow 0<\delta(s)$. 

Based on Figure \ref{fig:fig_sigma}, one assumes that $\sigma_{tot}^{pp}(s)$ achieves its minimum value at $s_{c1}$ and $\sigma_{tot}^{p\bar{p}}(s)$ at $s_{c2}$, being $s_{c1}<s_{c2}$. Considering the energy range $s_{c1}<s\rightarrow s_{c2}$, then $\sigma_{tot}^{p\bar{p}}(s)$ has been going toward its minimum value, while $\sigma_{tot}^{pp}(s)$ is growing as the energy rise (its minimum occurred at $s_{c1}$). At $s=s_{c2}$, on the other hand, one has that $\sigma_{tot}^{p\bar{p}}(s)$ achieve its minimum value, then $\delta(s)$ reaches its maximum in this energy range. Thus, one should expect the emergence of the visible shoulder in the region of the minimum of the total cross sections, as shown in Figure \ref{fig:fig_1}.   

It should be pointed out that if the minimum for both, $\sigma_{tot}^{pp}(s)$ and $\sigma_{tot}^{p\bar{p}}(s)$, occurred at the same energy $\sqrt{s}$, then $\delta(s)=0$, and a dip would be observed in this energy range. Nevertheless, it seems that it does not occur, as observed in Figure \ref{fig:fig_sigma}. 

\subsection{Total Cross Section and Entropy}

One supposes here that $\sigma_{tot}(s)$ can be divided into a finite number of non-interacting disjoint 2D cells, each one containing a $q\bar{q}$-pair ($u\bar{u}$ and $d\bar{d}$). Moreover, one neglects the spin since one supposes it does not change in the cell. 

Using $\delta(s)$, and assuming a classical view for the total cross section, the main point here is to furnish a measure of the entropy associated with the rise of $\sigma_{tot}(s)$. First of all, notice the result (\ref{eq:asym_2}) can be written as
\begin{eqnarray}\label{eq:alpha_2}
	\delta(s)= \ln\sigma_{tot}^{p\bar{p}}(s) - \ln\sigma_{tot}^{pp}(s).
\end{eqnarray}

Up to the Coulomb-Nuclear interference (CNI), the elastic scattering is governed by the electromagnetic and the strong interaction. In the near-forward direction, $t\approx 0$, both interactions contribute in a similar way to the interference effects measured, resulting in the high values observed to $\sigma_{tot}(s)$. However, one should stress that the $p\bar{p}$ total cross section, from CNI up to is minimum, is mainly due to baryon annihilation processes, which has higher total cross section, whereas the $pp$ total cross section is weakly energy-dependent in this energy range. 

As the collision energy rises above the CNI, one assumes the effective area represented by the (classical) total cross section can be divided into a finite number of identical non-interacting disjoint cells, each one containing a single $q\bar{q}$-pair subject to a confinement potential, as stated above. Consequently, each cell is viewed as possessing a null topological (color) charge, representing the experimental fact that the $q\bar{q}$-pair is a color singlet. Thus, one can write 
\begin{eqnarray}\label{eq:alpha_3}
	\delta(s)= \ln\left(\frac{\sigma_{tot}^{p\bar{p}}(s)}{a^2} a^2\right) - \ln\left(\frac{\sigma_{tot}^{pp}(s)}{b^2}b^2\right)=\ln\frac{\sigma_{tot}^{p\bar{p}}(s)}{a^2} - \ln\frac{\sigma_{tot}^{pp}(s)}{b^2}+\ln\left(\frac{a}{b}\right)^2,
\end{eqnarray}

\noindent where $a^2,b^2>0$ are the size of each cell in $\sigma_{tot}^{p\bar{p}}(s)$ and $\sigma_{tot}^{pp}(s)$, respectively. The interest here is the region near the minimum of the total cross sections, where the system is near the equilibrium (quasi-static). Thus, one can consider $a^2\approx b^2=c^2$. Therefore, near the minimum of the total cross section ($\ln (a/b)^2\approx 0$)
\begin{eqnarray}\label{eq:alpha_4}
	\delta(s)\approx \ln\frac{\sigma_{tot}^{p\bar{p}}(s)}{c^2} - \ln\frac{\sigma_{tot}^{pp}(s)}{c^2},
\end{eqnarray}

In this situation, one possible interpretation for each term in the r.h.s. of (\ref{eq:alpha_4}) is: they represent the entropy given by the presence of the cell of size $c^2$ (containing the $q\bar{q}$-pair) in the total cross sections. 

Near the minimum of $\sigma_{tot}(s)$, one can drop the scattering process, keeping the size of the cell as the same $pp$ and $p\bar{p}$. Then, one writes the entropy for the total cross section as
\begin{eqnarray}\label{eq:ent_1}
	S(s)=\ln \frac{\sigma_{tot}(s)}{c^2}.
\end{eqnarray}

The above definition is of statistical order since one can understand the calculation of $\sigma_{tot}(s)/c^2$ as the probability of finding a specific physical situation in the system. In thermodynamic terms, on the other hand, if the nature of the particles constituting the system and the interactions between them are fully known, then all the possible states $\Omega(s)$ of the system can be computed. If, near the equilibrium, these states are equally likely, then $\sigma_{tot}(s)/c^2\propto \Omega(s)$.

The total cross section in the classical picture is spherically symmetric and  $\sigma(s)\propto R^2(s)$, where $R(s)$ is the hadron radius. Of course, the size of the cell cannot be constant during the collision process. As the collision energy grows, the spatial separation between the quark and antiquark in the cell should decrease up to the non-confinement, near the QGP regime. However, near the minimum value of the total cross section, one can suppose the size of the cell is almost constant since $\sigma_{tot}(s)$ presents a small variation in this energy range (see Fig. \ref{fig:fig_sigma}). Accordingly, one can neglect small fluctuations in the volume of both the cell and hadron near the minimum. Taking the distance in the $q\bar{q}$-pair at the transition energy $s_c$ as $r_c$ as well as the hadron radius as $R(s_c)=R_c$, then the entropy (\ref{eq:ent_1}) at the minimum of $\sigma_{tot}(s)$ can be written as
\begin{eqnarray}\label{eq:ent_3}
	S(R_c)\approx 2\ln \frac{R_c}{r_c}.
\end{eqnarray}

The shape of the cell, square or circular, does not constitute a real problem since the 2D-disc and the unit square in $\mathbb{R}^2$ are homeomorphic. A theorem due to Prigogine \cite{I.Prigogine.Bull.Acad.Roy.Belg.Cl.Sci.31.600.1945} asserts that for steady states sufficiently close to equilibrium, the entropy production reaches its minimum. Then, at the minimum of $\sigma_{tot}(s)$, one gets the minimum value for the entropy in the system. 


\section{\label{bkt}BKT Phase Transition}

The study of phase transitions is one of the cornerstones of solid-state physics, relying on a well-defined ground. An important theorem due to Mermin and Wagner \cite{N.D.Mermin.H.Wagner.Phys.Rev.Lett.17.1133.1966} and  Hohenberg \cite{P.C.Hohenberg.Phys.Rev.158.383.1967} in statistical physics, and Coleman \cite{S.Coleman.Commun.Math.Phys.31.259.1973} in quantum field theory prevents the existence of ordered phases in the 2D system. In other words, in systems with finite temperature and with sufficiently short-range interaction, the Mermin-Wagner-Hohenberg-Coleman theorem states that it is impossible to have a phase transition from ordered to disordered system accompanied by a spontaneous symmetry breaking. Thus, long-range fluctuations are permitted and favored since their occurrence contributes to the increase of entropy. 

As well-known, the BKT phase transition occurs without spontaneous symmetry breaking \cite{V.L.Berezinskii.Zh.Eksp.Teor.Fiz.59.907.1970,J.M.Kosterlitz.D.J.Thouless.J.Phys.C6.1181.1973}. In the XY-model, at low temperatures, the presence of nontrivial vortex configuration is suppressed due to the binding effect caused by the vortex-antivortex interaction. At high temperatures, above some critical value, the binding effect can be neglected, and the nontrivial vortex becomes free. Consequently, one can estimate the critical temperature using the Helmholtz free energy.

In the BKT, the physical meaning of the phase transition is quite simple. For temperatures below the transition, the free energy is positive, and the single vortex is suppressed, resulting in the lattice behaving as a dilute gas of vortex-antivortex pairs. This is the so-called binding of vortex-antivortex pairs. Above the transition temperature, the free energy is negative, resulting in that entropy rise is favored by contributions coming from the free vortices. Then, the vortex-antivortex pairs are unbinding and the system becomes a vortex plasma. 

The internal energy and the entropy in the X-model possess the same logarithmic behavior, being that the key-point for the occurrence of the effect since this dependence provokes the competition between the energy $E$ and $S$ by the Helmholtz free energy. The dominance of one of these both quantities depends on the system is below or above the transition temperature.

As well-known, the internal energy of the hadron cannot be obtained from the first principles of thermodynamics or even from QCD. In general, the internal energy $U$ has several constraints, being the sum of the kinetic, potential, and chemical terms (among others). On the other hand, the strong interaction is the dominant energy in the $q\bar{q}$-pair. Thus, one can neglect the kinetic, chemical, and action terms in the system. 

The use of the potential to mimic the internal energy is not new in physics, remounting to the original work of Bohm \cite{D.Bohm.Phys.Rev.85.166.1952.ibid.180.1952}. Despite its naivety, the potential approach seems to agree, at a qualitative level, to the thermodynamics of $T$-matrix formalism \cite{Sh.F.Y.Liu_R.Rapp.arXiv:1501.07892.2015}. Also, the Bohm quantum potential has been used to mimic the internal energy of a quantum system, giving insight into its role in stationary states \cite{G.Dennis.M.A.de.Gosson.B.J.Hiley.Phys.Lett.A378.2363.2014,G.Dennis.M.A.de.Gosson.B.J.Hiley.Phys.Lett.A379.1224.2015}.

Differently from the original XY-model, the size of the cell as well the total cross section changes according to the collision energy. Furthermore, one may have a $q\bar{q}$-pair creation due to the rise of the collision energy. For simplicity, one assumes a constant number of pairs in the model. Then, one assumes the confinement potential $V(r)$ and the hadron internal energy $U(r)$ can be connected through the integration
\begin{eqnarray}\label{eq:int_ener}
	U(r)=\bar{m}\int_{r_0}^r V(r')dr',
\end{eqnarray}

\noindent where $r_0<r$ and $\bar{m}$ is the mean of the reduced mass $\mu(q\bar{q})$ of the different $q\bar{q}$-pairs: $u\bar{u}$ or $d\bar{d}$. Also, $r_0$ is the scale separating the confinement from the non-confinement regime of QCD. Notice the maximum of the spatial separation between the constituents of the pair is given at the end of CNI, where $r=R$, the hadron radius.

For the sake of simplicity, one reduces the focus of this work on two potentials: the Quigg-Rosner potential \cite{C.Quigg.J.L.Rosner.Phys.Lett.B71.153.1977} and the Cornell potential \cite{e_eichten_Phys_Rev_Lett_34_369_1975,e_eichten_Phys.Rev.D17.3090.1978,e_eichten_Phys.Rev.D21.203.1980}. These two potentials are power-law potentials where mass dependence of energy levels and distance scales follows elementary re-scaling operations \cite{C.Quigg.J.L.Rosner.Phys.Rep.56(4).167.1979}. Moreover, the virial theorem connects the kinetic $K$, potential and total bound-state $E(r)$ energies \cite{C.Quigg.J.L.Rosner.Phys.Rep.56(4).167.1979}
\begin{eqnarray}
	\langle K(r) \rangle = E(r) - \langle V(r) \rangle =\langle \frac{r}{2}\frac{d V(r)}{dr} \rangle.
\end{eqnarray}

Considering the above discussion, one can say these two potentials cover a wide class of physical interest. Then, the approach developed here is based on the main confinement potentials, and both agree with experimental results considering their limits of validity.

\subsection{Quigg-Rosner Confinement Potential}

There are several ways to confine the $q\bar{q}$-pairs \cite{e_eichten_Phys_Rev_Lett_34_369_1975,e_eichten_Phys.Rev.D17.3090.1978,e_eichten_Phys.Rev.D21.203.1980,a.martin.phys.lett.b93.338.1980,a.martin.phys.lett.b100.511.1988,a.martin.phys.lett.b21.561.1980,x.song.z.phys.c34.223.1987,d.b.lichtenberg.z.phys.c41.615.1989}. In particular, the Quigg-Rosner confinement potential is written as \cite{C.Quigg.J.L.Rosner.Phys.Lett.B71.153.1977}
\begin{eqnarray}\label{eq:conf_1}
	V(r)=\gamma \ln \frac{r}{r_0},
\end{eqnarray}

\noindent where $\gamma\approx 0.75$ GeV. This non-relativistic potential was used to reproduce some features of charmonium giving level spacing independent of the quark masses \cite{C.Quigg.J.L.Rosner.Phys.Lett.B71.153.1977}.  One can define the scale \cite{C.Quigg.J.L.Rosner.Phys.Rep.56(4).167.1979}
\begin{eqnarray}\label{eq:rc}
	r_{q\bar{q}}=\frac{1}{\mu(q\bar{q})},
\end{eqnarray}

\noindent as the confinement scale for each species of $q\bar{q}$ in the hadron. 
Then, one can write the confinement potential (\ref{eq:conf_1}) as the sum of the confinement potential due to $u\bar{u}$- and $d\bar{d}$-pairs
\begin{eqnarray}\label{eq:conf_2}
	V(r)=\gamma \left(\ln \frac{r}{r_{u\bar{u}}}+\ln \frac{r}{r_{d\bar{d}}}\right)=2\gamma \ln\frac{r}{r_0},
\end{eqnarray}

\noindent where one assumes that $\gamma$ is the same for both pairs, and that the confinement scale $r_0=\sqrt{r_{u\bar{u}}r_{d\bar{d}}}\approx 0.6$ fm, is the geometric mean of each individual confinement scale. It is important to stress that at the minimum of the total cross section, $r_c>r_0$.

Replacing the confinement potential (\ref{eq:conf_2}) into integral of (\ref{eq:int_ener}), one obtains
\begin{eqnarray}\label{eq:int_ener1}
	U(r)=2\bar{m}\gamma\left[r_0+r\left(\ln \frac{R_c}{r_0}-1\right)\right]- 2\bar{m}\gamma r \ln \frac{R_c}{r}.
\end{eqnarray}

From the experimental data available for $pp$ and $p\bar{p}$, the minimum for $\sigma_{tot}(s)$ is settled around $36\sim 42$ mb. Using $\sigma_{tot}(s)\approx 4\pi R^2$, then one obtains $R_c\approx 1.7\sim 1.9$ fm ($r_0<r_c\leq R_c$) at $s_c$. Without loss of generality, one supposes $r_c=nR_c$, where $n\leq 1$. Thus, for $r=r_c=nR_c$, the internal energy is written as (using $\ln y = \ln (yx/x)$, $y,x> 0$)
\begin{eqnarray}\label{eq:int_ener2}
	U(r_c)=2\bar{m}\gamma\left[r_0+nR_c\left(\ln \frac{r_c}{r_0}+\ln n-1\right)\right]+ 2\bar{m}\gamma nR_c \ln \frac{R_c}{r_c}.
\end{eqnarray}

The term $E_0=2\bar{m}\gamma[r_0+nR_c(\ln(R_c/r_0)+\ln n-1)]$ corresponds to the ground state energy at $s_c$. The ground state, defined here, means that the occupation of the energy states suffers an inversion, and the $q\bar{q}$-pair in the cell achieves the lower-energy state. Setting $E_0=0$, then one can obtain the value of $r_c$ according to $R_c$ and $r_0$. Taking $r_0\approx 0.6$ fm from the above considerations and assuming $R_c\approx 1.7$ fm at the minimum $\sigma_{tot}(s)$, one gets that for $n\approx 0.78$. Then, the spatial separation for the $q\bar{q}$-pair at the minimum of $\sigma_{tot}(s)$ is around $r_c\approx 1.33$ fm.

The Helmholtz free energy can give the transition temperature $T_c$ at $s_c$, following the scenario introduced above. The internal energy $U$ given by (\ref{eq:int_ener1}) and entropy (\ref{eq:ent_3}) has the same logarithmic dependence, competing by the free energy in the system. Then, the critical temperature is given by
\begin{eqnarray}\label{eq:temp_crit_1}
	T_c=m\gamma nR_c.
\end{eqnarray}

As stated above, the radius $R_c$ at $s_c$ is about $1.7\sim 1.9$ fm, and the mean mass is $\bar{m}\approx 1.8$ MeV, which result in the following critical temperature at the minimum of $\sigma_{tot}(s)$ 
\begin{eqnarray}\label{eq:temp_crit_2}
	T_c\approx 0.1 ~ \mathrm{MeV},
\end{eqnarray} 

\noindent corresponding to a temperature three orders of magnitude less than the Hagedorn temperature ($150\sim 200$ MeV), representing the approximate temperature where the $pp$ and $p\bar{p}$ total cross sections achieve their minimum value. 

\subsection{Cornell Confinement Potential}

The Cornell confinement potential is written as \cite{e_eichten_Phys_Rev_Lett_34_369_1975,e_eichten_Phys.Rev.D17.3090.1978,e_eichten_Phys.Rev.D21.203.1980}
\begin{eqnarray}\label{eq:cornell}
	V(r)=-\frac{4}{3}\frac{\alpha_s(r)}{r}+\sigma r,
\end{eqnarray}

\noindent where $\sigma$ is the string tension and $\alpha_s(r)$ is the running coupling constant of QCD. There is no general definition for $\alpha_s(r)$ in the confinement phase of QCD since the perturbative approach developed for the non-confinement cannot be used in that phase. This situation leads to different definitions for the running coupling as, for example, the approaches due to Richardson \cite{J.L.Richardson.Phys.Lett.B82.272.1979} and Brodsky-de T\'eramond-Deur (BdTD) \cite{S.J.Brodsky.G.F.deTeramond.A.Deur.Phys.Rev.D81.096010.2010}.

\begin{figure}
	\centering{
		\includegraphics[scale=0.35]{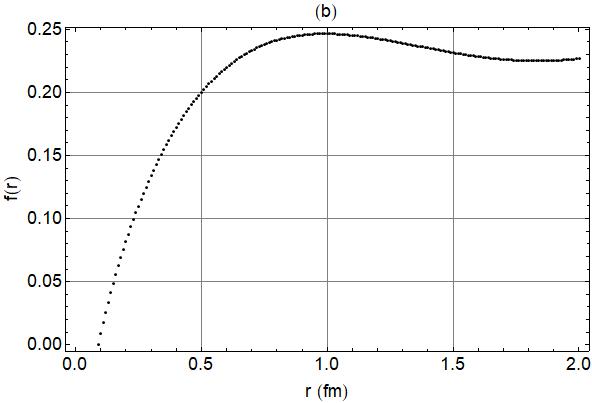}
		\includegraphics[scale=0.35]{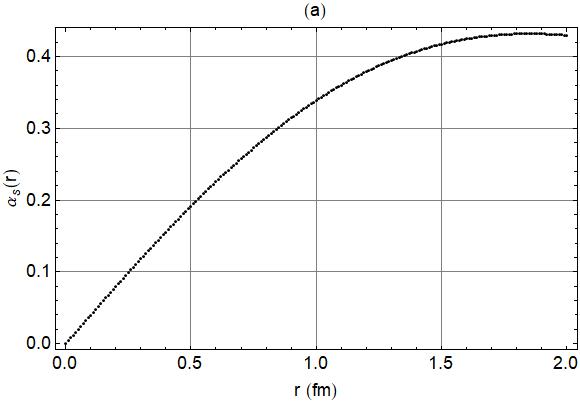}
		\caption{\label{fig:fig_num}Panel (b) shows the behavior of $f(r)$ given by (\ref{eq:fr}) using the running coupling as defined in \cite{S.J.Brodsky.G.F.deTeramond.A.Deur.Phys.Rev.D81.096010.2010}. In panel (b), the Fourier-sine transform of $\alpha_s(Q)$ given by (\ref{eq:bdtd_1}).}}
\end{figure}

Replacing (\ref{eq:cornell}) into (\ref{eq:int_ener}) and performing an integration by parts, one has
\begin{eqnarray}\label{eq:int_ener_cornell}
	U(r)=-\frac{4\bar{m}}{3}\left[\alpha_s(r)\ln r - \alpha_s(r_0)\ln r_0 + f(r)\right]+\frac{\sigma\bar{m}}{2}(r^2-r_0^2),
\end{eqnarray}

\noindent where 
\begin{eqnarray}\label{eq:fr}
	f(r)=-\int_{r_0}^r\alpha_s'(r')\ln r'dr',
\end{eqnarray} 

\noindent and $\alpha_s'(r)=d\alpha_s(r)/dr$. The function $f(r)$ depends on the definition of $\alpha_s(r)$ in the confinement phase of QCD. As an example, one concentrates the attention only on the BdTD approach, which is written as \cite{S.J.Brodsky.G.F.deTeramond.A.Deur.Phys.Rev.D81.096010.2010}
\begin{eqnarray}\label{eq:bdtd_1}
	\alpha_s(Q)=e^{-\frac{Q^2}{4k^2}},
\end{eqnarray}  
\noindent where $k$ is the mass scale. The numerical integration of (\ref{eq:fr}) is performed using the Fourier-sine transform of (\ref{eq:bdtd_1}), resulting in the running coupling in terms of $r$. Figure \ref{fig:fig_num}(a) shows the result for $f(r)$ while panel (b) exhibits the behavior of the Fourier-sine transform of (\ref{eq:bdtd_1}), $\alpha_s(r)$. Notice that the use of the mnemonic rule from QM, $Q\propto1/r$, will result, within this specific example, in a similar pattern for $\alpha_s(r)$. Based on the numerical results, one assumes the function $f(r)$ does not diverge for $r=r_c$.

Observe that one can write (\ref{eq:int_ener_cornell}) as
\begin{eqnarray}\label{eq:int_ener_cornell_2}
	U(r)=-\frac{4\bar{m}}{3}\left[\ln \left(\frac{R_c}{r_0^{\alpha_s(r_0)}}\right) + f(r)\right]+\frac{\sigma\bar{m}}{2}(r^2-r_0^2)+\frac{4\bar{m}}{3}\ln\left(\frac{R_c}{r^{\alpha_s(r)}}\right).
\end{eqnarray} 

Of course, the running coupling at $r_c$ is finite, and it can be lower or greater than 1. If it is greater than 1, then $\alpha_s(r_c)=1+e(r_c)$, where $e(r_c)$ is the excess of $\alpha_s(r_c)$. If it is lower than 1, then $\alpha_s(r_c)=1-l(r_c)$ where $l(r_c)$ is the lack of $\alpha_s(r_c)$. Without any loss, one supposes $\alpha_s(r_c)<1$ and, then, the last term on the r.h.s. of (\ref{eq:int_ener_cornell_2}) can be written as (at $r=r_c$)
\begin{eqnarray}
	\ln\left(\frac{R_c}{r_c^{\alpha_s(r_c)}}\right)=\ln\left(\frac{R_c}{r_c}\right)+\ln\left(\frac{1}{r_c^{-l(r_c)}}\right).
\end{eqnarray} 

Returning this term within (\ref{eq:int_ener_cornell_2}) and reorganizing the result, one can, analogously to the internal energy obtained from the Quigg-Rosner potential, define the ground state energy $E_0=0$ at $r_c$ 
\begin{eqnarray}
	E_0=-\frac{4\bar{m}}{3}\left[\ln \left(\frac{r_0^{\alpha_s(r_0)}}{R_c r_c^{-l(r_c)}}\right)+ f(r_c)\right]+\frac{\sigma\bar{m}}{2}(r_c^2-r_0^2)=0.
\end{eqnarray}

The temperature at the phase transition can be calculated assuming, again, that near the phase transition, the Helmholtz free energy holds, resulting
\begin{eqnarray}
	T_c=\frac{2}{3}\bar{m}\approx 1.2 ~~ \mathrm{MeV},
\end{eqnarray} 

\noindent which is $\sim 10$ times greater than the temperature obtained using the Quigg-Rosner confinement potential. The definition of $T_c$ from the Cornell confinement potential does not depend on $\alpha_s(r_c)$, contrarily to the ground state energy.

The conclusion from the above results allows one to assert that either the Quigg-Rosner as well as Cornell confinement potential, there is a topological phase transition occurring at $T_c\approx 0.1\sim 1.2$ MeV. 

A possible explanation for the emergence of the topological phase transition may be the presence of the so-called quarkyonic matter, introduced some years ago \cite{L.McLerran.R.D.Pisarski.Nucl.Phys.A796.83.2007,L.YA.Glozman.R.F.Wagenbrunn.Mod.Phys.Lett.A23.2385.2008}. In this picture, the chiral symmetry is restored even in face of confinement matter (hadrons). Thus, the quark matter in the hadron undergoes the phase transition obtained here, whose result is the restoration of the chiral symmetry above $T_c$. Therefore, above $T_c$ but below the Hagedorn temperature, the emergence of a free quark state is allowed \cite{L.McLerran.R.D.Pisarski.Nucl.Phys.A796.83.2007,L.YA.Glozman.R.F.Wagenbrunn.Mod.Phys.Lett.A23.2385.2008}, resulting in an effect quite similar to the emergence of the free vortex in the XY-model. Indeed, it was shown that the chiral symmetry can be restored but the matter still remains confined \cite{L.YA.Glozman.R.F.Wagenbrunn.Mod.Phys.Lett.A23.2385.2008}. 

\section{\label{fr}Discussion}

One cannot disregard the entropy in the study of any physical system. Assuming the classical view of the total cross section, one divides $\sigma_{tot}(s)$ in a finite number of non-interacting disjoint 2D cells, each one containing a $q\bar{q}$-pair. Analogously to the XY-model, one associate an entropy to the area represented by the logarithm of the ratio of the total cross section to the area of the cell. Using the Quigg-Rosner and the Cornell confinement potentials, one calculates the internal energy of the hadron. It was shown here that both entropy and internal energy has the same logarithmic dependence on the spatial distance between the constituents of the cell. Then, near the minimum of the total cross section, one supposes a small (volume) variation, which allows the use of the Helmholtz free energy to calculate the transition temperature at the minimum of $\sigma_{tot}(s)$.

The use of the Quigg-Rosner confinement potential was able to furnish the desired logarithmic dependence on the spatial pair separation. The Cornell confinement potential, which has a good agreement with experimental data \cite{M.G.Olsson.S.Vesell.K.Williams.Phys.Rev.D51.5079.1995,D.Ebert.V.O.Galkin.R.N.Faustov.Phys.Rev.D57.5663.1998,E.J. Eichten.C.Quigg.Phys.Rev.D49.5845.1994} for the spectrum of light and heavy mesons, also has furnished the same behavior. Certainly, due to this good agreement with the experimental data, it is possible to suppose that the transition temperature $T_c$ is near 1.2 MeV than the value 0.1 MeV, obtained within the Quigg-Rosner approach. 

It is important to stress this transition temperature does not correspond to the Hagedorn temperature since, at this collision energy, the hadron is far from the transition from the confinement to the non-confinement phase of QCD. Thus, the transition temperature obtained corresponds to the one where here is supposed a topological phase transition takes place. A possible explanation is based on the quarkyonic matter \cite{L.McLerran.R.D.Pisarski.Nucl.Phys.A796.83.2007,L.YA.Glozman.R.F.Wagenbrunn.Mod.Phys.Lett.A23.2385.2008}, where the presence of free quarks is allowed in the confinement phase of QCD. This approach leads to a direct identification between the appearance of the free quark and the emergence of the free vortex in the XY-model. In the work of McLerran and Pisarski \cite{L.McLerran.R.D.Pisarski.Nucl.Phys.A796.83.2007}, the main idea is based on qualitative arguments for large $N_c$, stating that at large chemical potential and low temperature (as obtained here), the existence of a chirally symmetric phase of confining quark is possible. On the other hand, Glozman and Wagenbrunn \cite{L.YA.Glozman.R.F.Wagenbrunn.Mod.Phys.Lett.A23.2385.2008} show that, with a finite density and low temperature, the same effect can be found. Thus, the improvement of the present model should include information on the finite density of the system, for example, which will be performed elsewhere.

The approach for entropy presented here is, of course, not performed from the first principles of thermodynamics, which is not obvious at all. Thus, this is a clear weakness of the approach. Also, the use of a potential to mimic the internal energy is not new in physics, but it evidently excludes all the other energy manifestation forms in the hadron. However, despite the naivety of the presented approach, it corroborates recent theoretical strategies indicating the presence of a chirally symmetric phase (free quarks) in the confinement phase of QCD, with remarkable similarity with the BKT approach for the XY-model.

In a more realistic scenario, the creation of virtual $q\bar{q}$-pairs may occur in the off mass shell regime. The back-reaction of the virtual pair creation may act to raise the total cross section. In other words, the increasing number of virtual cells may lead to the growth of entropy, implying the rise of the total cross section.


\section*{Acknowledge}

SDC thanks to UFSCar by the financial support.

\section*{References}

\end{document}